\documentclass[aps,ams,amsmath,prx,twocolumn,superscriptaddress]{revtex4-1}
\usepackage{graphics}
\usepackage{graphicx}
\usepackage{epsfig}
\usepackage{amsmath}
\usepackage{amssymb}
\usepackage{amsfonts}
\usepackage{bm}
\usepackage{color}
\usepackage{bbm}
\usepackage{hyperref}
 \usepackage[normalem]{ulem}
\usepackage{soul}
\DeclareMathOperator{\Det}{Det}

\begin{document}

\title{Hydrodynamic thermoelectric transport in Corbino geometry}

\author{Songci Li}
\affiliation{Department of Physics, University of Wisconsin-Madison, Madison, Wisconsin 53706, USA}

\author{Alex Levchenko}
\affiliation{Department of Physics, University of Wisconsin-Madison, Madison, Wisconsin 53706, USA}

\author{A. V. Andreev}
\affiliation{Department of Physics, University of Washington, Seattle, Washington 98195, USA}
\affiliation{Skolkovo  Institute of  Science  and  Technology,  Moscow,  143026,  Russia}
\affiliation{L. D. Landau Institute for Theoretical Physics, Moscow, 119334 Russia}

\date{March 8, 2022}

\begin{abstract}
We study hydrodynamic electron transport in Corbino graphene devices. Due to the irrotational character of the flow, the forces exerted on the electron liquid are expelled from the bulk. We show that in the absence of Galilean invariance, force expulsion produces qualitatively new features in thermoelectric transport: (i) it results in drops of both voltage and temperature at the system boundaries and (ii) in conductance measurements in pristine systems, the electric field is not expelled from the bulk.  We obtain thermoelectric coefficients of the system in the entire crossover region between charge neutrality and high electron density regime. The thermal conductance exhibits a sensitive Lorentzian dependence on the electron density. The width of the Lorentzian is determined by the fluid viscosity. This enables determination of the viscosity of electron liquid near charge neutrality from purely thermal transport measurements. In general, the thermoelectric response is anomalous: it violates the Matthiessen's rule, the Wiedemann-Franz law, and the Mott relation.
\end{abstract}

\maketitle

\section{Introduction}

Hydrodynamic electron transport in graphene devices has been the subject of active experimental \cite{Bandurin-1,Crossno,Ghahari,Morpurgo,Kumar,Bandurin-2,Berdyugin,Hone,Brar} and theoretical research \cite{ Muller,MFS,FSMS,Kashuba,MSF,Aleiner,Phan-Song,NGTSM,Xie-Foster,Lucas,Alekseev,Principi-2DM,LLA,NG,AVA,LAL} over the past few years, see reviews \cite{NGMS,Lucas-Fong,ALJS} and references therein. In most electron systems the hydrodynamic flow corresponds to the flow of charge. The peculiarity of electron hydrodynamics in graphene is that at charge neutrality, the hydrodynamic flow carries no charge and corresponds to pure heat transport~\cite{Aleiner,Phan-Song,NGTSM}. The accurate  control of electron density in graphene devices enables investigation of the full crossover between the entropy-dominated and charge-dominated regimes of hydrodynamic transport.

In large graphene monolayer samples, Refs.~\cite{Crossno,Ghahari} investigated this crossover and elucidated the anomalous thermoelectric response in a Dirac fluid. In this case, the crossover width is determined by the bulk inhomogeneities of the device~\cite{Lucas,Principi-2DM,LLA}. Recently experimental \cite{Fuhrer,Geim,Bykov,Dietsche,Hakonen}
and theoretical \cite{Tomadin,Holder,Shavit,Principi} efforts focused on hydrodynamic electron transport in the Corbino geometry. The interest in this geometry is that even in a pristine system the hydrodynamic flow generates energy dissipation associated with viscous stresses. This enables determination of intrinsic dissipative properties of the electron liquid from transport measurements.

Another peculiarity of the Corbino setup is related to the purely potential character of the flow. In this case, in Galilean-invariant liquids the Bernoulli law holds despite the presence of dissipative stresses~\cite{Faber,Falkovich}. In linear transport, this corresponds to spatially uniform pressure in the liquid. In particular, for charged Galilean-invariant liquids, this manifests in expulsion of the electric field from the interior of the system \cite{Shavit}. This effect has been probed in recent local imaging experiments \cite{Ilani}. Furthermore, magnetometry and scanning probe techniques, in general, allow direct visualization of the viscosity-dominated electronic flow profile \cite{Sulpizio,Ku,Jenkins}. High-quality electron magnetotransport measurements in graphene Corbino devices have been also reported \cite{Dean}.

These advances motivate theoretical description of hydrodynamic electron transport in Corbino devices in the full crossover between the regimes of charge neutrality and high electron density. An important aspect of these systems is the absence of Galilean invariance of the electron liquid. Below we develop such a theory and describe thermoelectric response of the system as a function of electron density and temperature. We show that in the absence of Galilean invariance, uniformity of pressure and expulsion of force from the bulk leads to qualitatively new consequences. In Galilean-invariant liquids~\cite{Falkovich}, force expulsion corresponds to expulsion of the electric field from the bulk flow and produces a voltage jump at the system boundary. In the absence of Galilean invariance it produces discontinuities not only in voltage but also in temperature at the system boundary. This temperature jump may not be attributed to the Kapitza boundary resistance, which occurs at interfaces between liquid helium and solids \cite{Kapitza,Khalatnikov}, or, more generally, between two media with mismatched acoustic impedances at low temperatures \cite{Swartz-Pohl}. In particular, for a thermal resistance measurement in the present case, the temperature of the liquid is ether higher (for centripetal flow) or lower (for centrifugal flow) than the temperatures of both contacts. This is in striking contrast with the Kapitza resistance situation, where the heat flux across the boundary flows from the medium with higher temperature to that with lower temperature. This difference can be traced to the fact that entropy production in the present case occurs inside flowing liquid rather than the contacts. As a result, the positive-definite thermal resistance of the system cannot be written as a sum of positive-definite contributions of the boundaries.

\begin{figure*}[t!]
\includegraphics[width=0.3\linewidth]{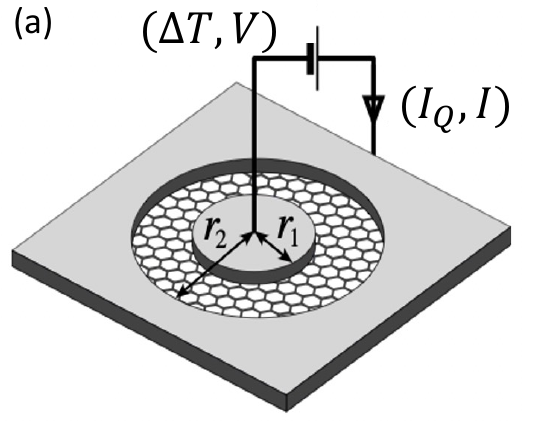}
\includegraphics[width=0.3\linewidth]{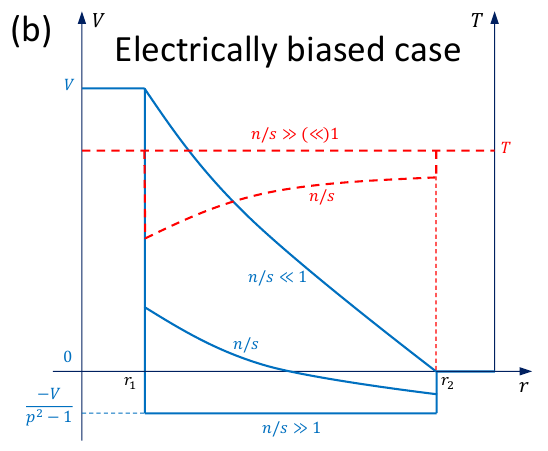}
\includegraphics[width=0.315\linewidth]{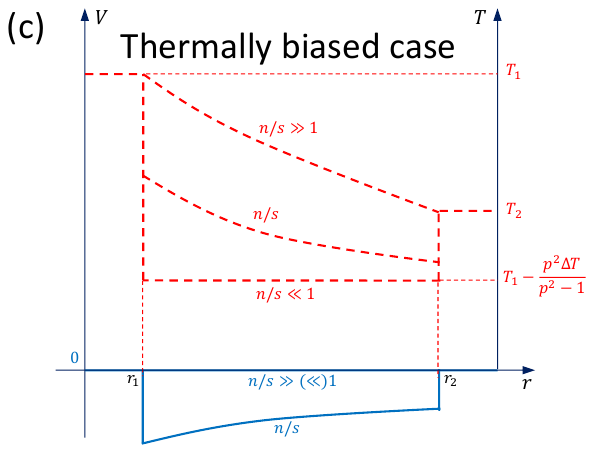}
\caption{(a) A schematic setup for the graphene Corbino device subject to a mixed thermoelectric bias with the temperature difference $\Delta T$ and voltage $V$ applied between the electrodes that generate heat current $I_Q$ and electric current $I$. Panels (b) and (c) show sketches for the spatial profile of the electric potential (blue) and temperature (red) at different densities for purely electrical and thermal bias, respectively. Voltage drop occurs entirely in the bulk at charge neutrality, and at the boundary for large density limit, and conversely for the temperature drop. Note the existence of spatial regions where temperature/potential of the liquid is lower than that of both contacts. }
\label{Fig-CVT}
\end{figure*}

The appearance of a temperature jump at the boundary leads to another qualitative difference with Galilean-invariant systems: in a linear conductance measurement, the electric field is no longer expelled from the flow, even in pristine systems. The reason is that in a general situation, force expulsion does not require vanishing of the electric field, only that the force due to the electric field must be compensated by the force caused by  the temperature gradient. Since the conductance is measured at zero temperature difference, the temperature drop at the system boundary must be compensated by temperature gradients in the bulk. Because of the force expulsion, this produces electric field in the bulk flow.

The thermoelectric properties of the systems arise from two modes of charge and entropy transport, namely, the hydrodynamic flow and transport relative to the liquid. The respective contributions to resistance have different dependence on the electron scattering time, one being proportional to it whereas the other is  inversely proportional. This signifies the breakdown of Matthiessen's rule in the hydrodynamic regime.  For the resulting thermoelectric response, we find strong violation of the Wiedemann-Franz law and enhanced Seebeck coefficient when compared to the usual Lorenz number and semiclassical Mott formula of single-particle transport, respectively.

\section{Hydrodynamic description}

We consider radial charge and heat transport in a Corbino device, which is contacted by the inner electrode of radius $r_1$ and the outer electrode of radius $r_2$, as illustrated in panel (a) of Fig.~\ref{Fig-CVT}. We assume that  small electric current $I$ and heat current $I_Q$ are induced in  the device by voltage $V$ and temperature difference $\Delta T$ between the electrodes. The hydrodynamic regime arises when the rate of momentum-conserving electron-electron collisions exceeds the rate of extrinsic processes leading to momentum and energy relaxation. In linear response, both the particle and the entropy currents are conserved. Denoting the net electron current by $I_n$ and the entropy current by $I_s$, we write the continuity equation of particle and entropy currents in the form
\begin{equation}\label{continuity}
\vec{x}u-\hat{\Upsilon}\vec{X}=\frac{\vec{I}}{2\pi r}, \quad \vec{I}=\left(\begin{array}{c}I_n \\ I_s\end{array}\right),
\end{equation}
where $u(r)$ is the radial hydrodynamic velocity, and $r\in[r_1,r_2]$ is the radial coordinate. Here we introduced the two-component column vector of particle and entropy currents, $\vec{I}$, the column vector of particle and entropy densities,  $n$ and $s$,  $\vec{x}^{\mathbb{T}}=(n,s)$ (with the superscript $\mathbb{T}$ indicating the transposition), and the corresponding column-vector  of thermodynamically conjugate forces $\vec{X}^{\mathbb{T}}(r)=(-e\mathcal{E},\nabla T)$ \cite{LL-V5}. In the latter, $e\mathcal{E}$ represents the electromotive force (EMF). The matrix $\hat{\Upsilon}$ characterizes the dissipative properties of the electron liquid. In the absence of Galilean invariance it is given by
\begin{equation}
\hat{\Upsilon}=\left(\begin{array}{cc}\sigma/e^2 & \gamma/T \\ \gamma/T & \kappa/ T\end{array}\right)
\end{equation}
and consists of the thermal conductivity $\kappa$, the intrinsic conductivity $\sigma$, and the thermoelectric coefficient $\gamma$, see Refs.~\cite{FSMS,MSF,Aleiner,Kashuba}. For Galilean-invariant liquids, we have $\sigma =\gamma =0$.
Equation \eqref{continuity} should be supplemented by  the Navier-Stokes equation, which for a steady-state linear response flow expresses local force balance. The radial force per unit area caused by the temperature gradient and EMF can be expressed in the column vector notations as $\vec{x}^{\mathbb{T}}\vec{X}$~\cite{LLA}, and for the radial flow we have
\begin{equation}\label{NS}
(\eta + \zeta) \hat{\Delta}u=\vec{x}^{\mathbb{T}}\vec{X},\qquad \hat{\Delta}=\frac{1}{r}\frac{d}{dr}\left(r\frac{d}{dr}\right)-\frac{1}{r^2}.
\end{equation}
Here $\eta$ and $\zeta$ are, respectively, the shear and bulk viscosities, which appear in the expression for the viscous stress tensor:
\begin{equation}
\Sigma_{ij}=\eta(\partial_iu_j+\partial_ju_i)+(\zeta-\eta)\delta_{ij}\partial_ku_k.
\end{equation}
For single-layer graphene, the bulk viscosity  $\zeta$ is expected to be negligible due to scale invariance of 2D electron systems with linear dispersion and Coulomb interactions~\cite{LL-V10,Son}.

Equations \eqref{continuity} and \eqref{NS} determine the spatial dependence of the flow velocity, electric field, and temperature gradients in the interior of the Corbino disk in terms of the particle and entropy currents $\vec{I}$.
Expressing the vector $\vec{X}$ of electromotive and thermal gradient forces in terms of the radial velocity and currents using Eq.~\eqref{continuity}, and substituting the result into the Navier-Stokes equation \eqref{NS}, we obtain the following:
\begin{equation}\label{NS-u}
\hat{\Delta}u-\frac{u}{l^2}=-\frac{1}{2\pi r (\eta +\zeta)}(\vec{x}^{\mathbb{T}}\hat{\Upsilon}^{-1}\vec{I}).
\end{equation}
Here we introduced the length scale $l$ defined by
\begin{equation}
l^{-2}=\frac{{\vec x}^{\mathbb{T}}\hat{\Upsilon}^{-1}\vec{x}}{\eta + \zeta} =\frac{s^2 \left[\frac{\sigma}{e^2}+\frac{n^2\kappa}{s^2T}-\frac{2n\gamma}{sT}\right]}{(\eta + \zeta)\left[\frac{\sigma}{e^2}\frac{\kappa}{T } - \frac{\gamma^2}{T^2} \right]}.
\end{equation}
The general solution of Eq. \eqref{NS-u} is given by  the sum of a linear combination of modified Bessel functions of the first and second kinds, $\mathrm{I}_1(r/l)$ and $\mathrm{K}_1(r/l)$, and the particular solution of the inhomogeneous equation, which has the form
\begin{equation}\label{u}
u(r)=\frac{1}{2\pi r}\frac{{\vec x}^{\mathbb{T}}\hat{\Upsilon}^{-1}\vec{I}}{{\vec x}^{\mathbb{T}}\hat{\Upsilon}^{-1}\vec{x}}.
\end{equation}
The latter describes the flow in the bulk of the Corbino disk, namely, at distances greater then $l$ away from the boundaries.  The exponentially decaying and growing solutions of a homogeneous equation, $\mathrm{K}_1 (r/l)$ and $\mathrm{I}_1(r/l)$, describe the deviations from the bulk flow \eqref{u} near the inner and outer boundaries, and contribute to the thermoelectric resistance of the contacts.

The thermoelectric resistance matrix  $\hat{\mathcal{R}}$ can be obtained by equating  the Joule heat,
\begin{equation}\label{P}
\mathcal{P}=\vec{\mathcal{I}}^{\mathbb{T}}\hat{\mathcal{R}}\vec{\mathcal{I}},\qquad \vec{\mathcal{I}}^{\mathbb{T}}=(I,I_Q),
\end{equation}
to the rate of energy dissipation in the bulk flow. Here $I=eI_n$ and $I_Q =TI_s$ are the electric and heat currents. We are interested in the bulk contribution to the thermoelectric resistance matrix. To this end, we neglect the deviations of the flow from the bulk flow \eqref{u}, which contribute to the contact resistance. The total resistance matrix is obtained by adding to it the resistance matrices of the contacts. Keeping in mind the divergenceless character of the bulk flow, we can express the latter in the form
\begin{equation}\label{dissipation}
\mathcal{P}=\frac{1}{2\eta}\int\sum_{ij}\Sigma^2_{ij}\,d^2r+\int \vec{X}^{\mathbb{T}}\hat{\Upsilon}\vec{X}\,d^2r.
\end{equation}
In this expression, the first term accounts for the viscous dissipation generated by the hydrodynamic transport mode [first term on the left-hand side of Eq.~\eqref{continuity}]. The second term describes the entropy production due to the transport in the relative mode [second term on the left-hand side of Eq.~\eqref{continuity}], i.e., charge and energy transport relative to the liquid.

We begin by considering the contribution of the relative transport mode to the dissipation rate. The EMF and temperature gradients corresponding to the bulk flow $u(r)$ in Eq. \eqref{u} can be obtained by multiplying  Eq.~\eqref{continuity} by a row vector $\vec{\chi}^{\mathbb{T}} = (s, -n)$ from the left. As expected, the result,
\begin{equation}\label{eq:X_bulk}
  \vec{X}(r) = - \frac{\vec{\chi}}{2 \pi r}  \, \frac{\vec{\chi}^{\mathbb{T}} \vec{I}}{\vec{\chi}^{\mathbb{T}} \hat{\Upsilon} \vec{\chi}},
\end{equation}
obeys the vanishing force density  condition ,  $\vec{x}^{\mathbb{T}}\vec{X} =0$, or more explicitly
\begin{equation}\label{eq:force_expulsion}
- n e \mathcal{E} + s \nabla T =0 .
\end{equation}
This corresponds to uniform pressure in the bulk~\cite{Faber,Falkovich}.
Substituting the expression \eqref{eq:X_bulk} into Eq.~\eqref{dissipation} we obtain the rate of energy dissipation due to the relative transport mode.

The viscous contribution to the dissipation rate in Eq.~\eqref{dissipation}  is evaluated by substituting the velocity \eqref{u} into the viscous stress tensor.   For a  radial flow, there are only two nonvanishing components of the stress tensor. In cylindrical coordinates, these are \cite{LL-V6}
\begin{equation}
\label{sigma}
\Sigma_{rr}=2\eta\frac{\partial u}{\partial r},\quad \Sigma_{\phi\phi}=2\eta\frac{u}{r}.
\end{equation}
After a simple calculation we find the dissipated power $\mathcal{P}$ in the form of Eq. \eqref{P} with the elements of the resistivity matrix given by
\begin{subequations}\label{R}
\begin{align}
\label{R-a}
&\mathcal{R}_{11}=\frac{1}{2\pi e^2}\left[\frac{2\eta}{s^2}\frac{\varkappa^2(r^{-2}_1-r^{-2}_2)}{(\varsigma+n\varkappa/s)^2}+\frac{\ln p}{\varsigma+n\varkappa/s}\right], \\
\label{R-b}
&\mathcal{R}_{22}=\frac{1}{2\pi T^2}\left[\frac{2\eta}{s^2}\frac{\varsigma^2(r^{-2}_1-r^{-2}_2)}{(\varsigma+n\varkappa/s)^2}+\frac{(n/s)^2\ln p}{\varsigma+n\varkappa/s}\right],\\
\label{R-c}
&\mathcal{R}_{12}=\frac{1}{2\pi eT}\left[\frac{2\eta}{s^2}\frac{\varsigma\varkappa(r^{-2}_1-r^{-2}_2)}{(\varsigma+n\varkappa/s)^2}-\frac{(n/s)\ln p}{\varsigma+n\varkappa/s}\right],
\end{align}
\end{subequations}
and $\mathcal{R}_{21}=\mathcal{R}_{12}$. Here we introduced two dimensionless quantities
\begin{equation}
\varsigma=\frac{\sigma}{e^2}-\frac{n\gamma}{sT}, \quad \varkappa=\frac{n\kappa}{sT}-\frac{\gamma}{T},
\end{equation}
and also aspect ratio of the disk $p=r_2/r_1>1$.

The first terms in square brackets in Eqs.~\eqref{R} represent the contributions to the thermoelectric resistance of the hydrodynamic transport mode. It is easy to see that they are proportional to the electron-electron relaxation time. This reflects the Gurzhi effect~\cite{Gurzhi,LWM1,LWM2} -- decrease of resistivity with increasing relaxation rate. The second terms in \eqref{R} correspond to the contribution of the relative transport mode and are inversely proportional to the relaxation time. The opposite scaling of these additive contributions to the resistance with the relaxation rate implies violation of Matthiesen's rule.

Note that although the resistance matrix can be written as a sum of two contributions, which depend on the inner and outer radii, these contributions are not positive-definite. Therefore, the resistance matrix cannot be interpreted as a sum of contact resistances. This reflects the fact that the entropy production occurs in the bulk of the flow. The integrated dissipation is given by a difference of functions evaluated at $r_1$ and $r_2$. Another point is that the contribution of the relative mode, namely, the logarithmic terms, are associated with the drop of voltage and temperature in the bulk. In contrast, the contribution of the hydrodynamic mode is related to the voltage and temperature drops at the contacts.

\section{Transport coefficients}

Since the dissipated power is also expressed as $\mathcal{P}=\vec{\mathcal{I}}^\mathbb{T} \vec{\mathcal{X}}$, where $\vec{\mathcal{X}}=(V,\Delta T/T)^{\mathbb{T}}$ are the applied voltage and temperature difference, the inverse matrix of $\hat{\mathcal{R}}$, $\hat{\mathcal{G}} \equiv \hat{\mathcal{R}}^{-1}$, relates the linear response of current $\vec{\mathcal{I}}$ to the applied voltages, i.e., $\vec{\mathcal{I}}=\hat{\mathcal{G}}\vec{\mathcal{X}}$. Therefore, $\hat{\mathcal{G}}$ is the macroscopic conductance matrix. These general results are applied below to common setups used in thermoelectric measurements.

\subsection{Thermal resistance} 

The thermal resistance $R_\text{th}$, is obtained by setting the electric current $I$ to zero. The computed dissipation $\mathcal{P}$ in Eq.~\eqref{P} is given by $\mathcal{P}=\mathcal{R}_{22}I^2_Q$. Employing thermodynamic relations, one finds the entropy production rate $\dot{\mathcal{S}}=I_Q\Delta T/T^2=R_\text{th}I^2_Q/T^2$, and the dissipated power $\mathcal{P}=T\dot{\mathcal{S}}=R_\text{th}I^2_Q/T$. Therefore, the thermal resistance is $R_\text{th}=T\mathcal{R}_{22}$. Using Eq.~\eqref{R-b}, we get
\begin{equation}\label{R-th}
R_{\text{th}}=\frac{1}{2\pi T}\left[\frac{2\eta}{s^2r^2_2}\frac{\varsigma^2(p^2-1)}{(\varsigma+n\varkappa/s)^2}+\frac{(n/s)^2\ln p}{\varsigma+n\varkappa/s}\right].
\end{equation}
At charge neutrality $n\to0$, the second term vanishes. We also note that the thermoelectric coefficient vanishes at the Dirac point, $\gamma\to0$, due to approximate particle-hole symmetry. Therefore, at the charge neutrality point, the thermal resistance is determined by the viscosity of the electron liquid, $R_{\text{th}}=A\eta/Ts^2$, where $A=(p^2-1)/\pi r^2_2$ is the geometric coefficient. For a graphene monolayer, $\eta\sim s\sim (T/v)^2$, up to an additional logarithmic renormalizations of viscosity \cite{Muller}. Therefore, we conclude that $R_{\text{th}}\propto 1/T^3$. In the opposite limit of high density $n\gg s$, the thermal resistance is dominated by the bulk term, which reduces to $R_{\text{th}}=\ln p/(2\pi\kappa)$.

To elucidate the physical origin of the viscous contribution to the thermal resistance, it is instructive to derive it from an alternative consideration. The radial viscous tresses arising in the liquid exert additional radial force on the contacts, which must be compensated by excess pressure. This pressure difference may be related to the voltage and temperature drop at the contacts by the thermodynamic identity $dP=nd\mu+sdT$~\cite{LL-V5}.
Let us focus, for the sake of clarity, on the charge neutrality point, $n=0$. In this case, we find from the force expulsion condition \eqref{eq:force_expulsion} that the temperature of the liquid  must be uniform  in the bulk, $T(r) =T_l$, whereas the pressure jumps at the boundary with the contacts are given by
\begin{equation}\label{boundary}
s (T_i - T_l)=\Sigma_{rr}(r_i),
\end{equation}
where $T_i$ is the temperature of the $i$-th contact. Using $u(r)$ from Eq. \eqref{u} at $I_n=0$ and calculating $\Sigma_{rr}$ from Eq. \eqref{sigma} at both boundaries, we obtain
\begin{equation}
T_l - T_i=- \frac{\eta I_s}{\pi s^2} \frac{1}{r^2_i}.
\end{equation}
The net temperature drop is consistent with  Eq. \eqref{R-th} for $n\to0$.
Note that the temperature of the liquid, $T_l$ is either higher or lower than the temperatures of both leads, depending on the direction of the heat flow. As explained above, this is a consequence of the bulk character of entropy production.
A similar consequence of force expulsion occurs in charge transport away from charge neutrality. The voltage drop between the contacts and the electron liquid has the same sign at both boundaries. This behavior is illustrated in Figs. \ref{Fig-CVT}(b) and \ref{Fig-CVT}(c),
showing voltage and temperature distributions for varying density at different biasing setups. These predictions may be tested via high-resolution thermal imaging and scanning gate microscopies \cite{Zeldov-Nature,Zeldov-Science}.

\begin{figure*}[t!]
 \includegraphics[width=0.245\linewidth]{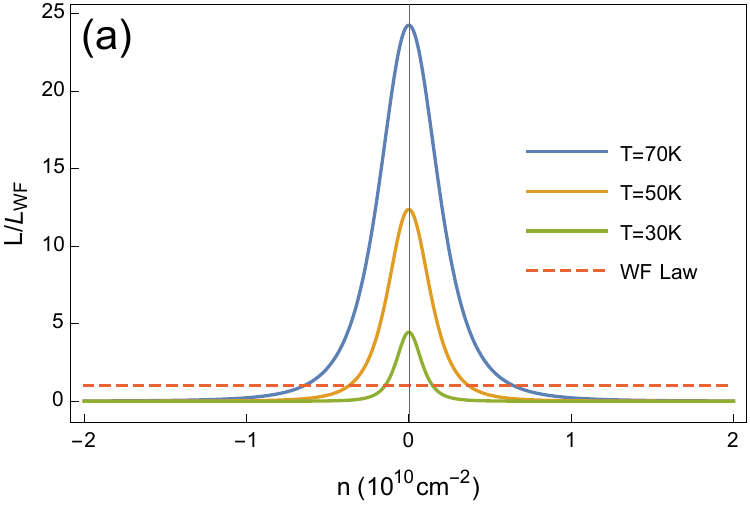}
\includegraphics[width=0.245\linewidth]{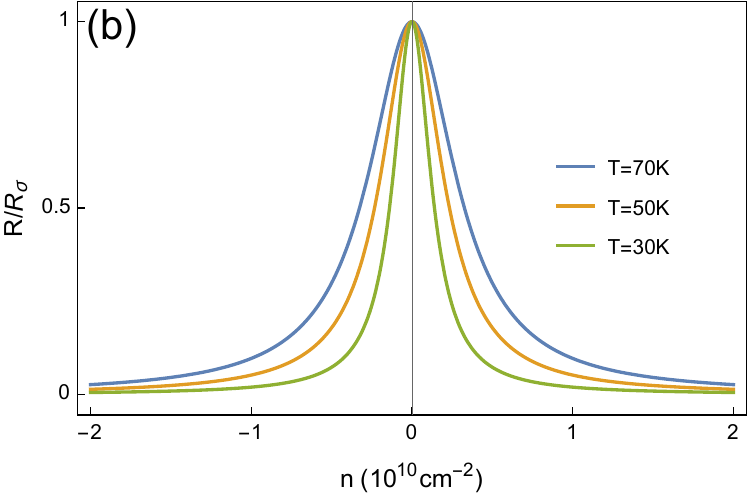}
\includegraphics[width=0.245\linewidth]{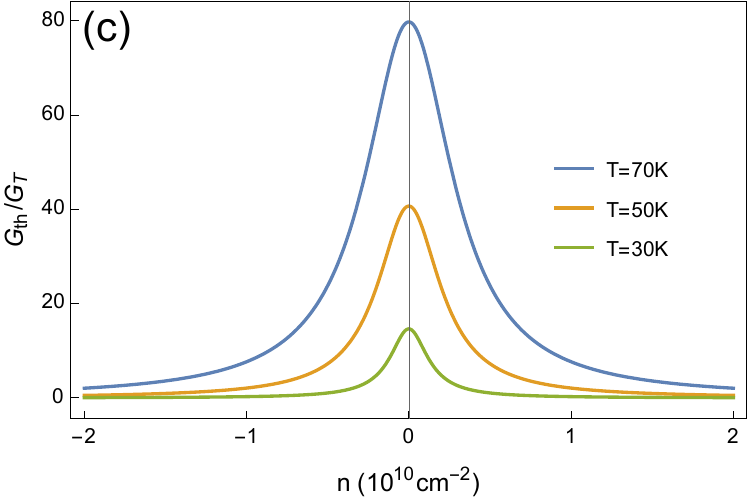}
\includegraphics[width=0.245\linewidth]{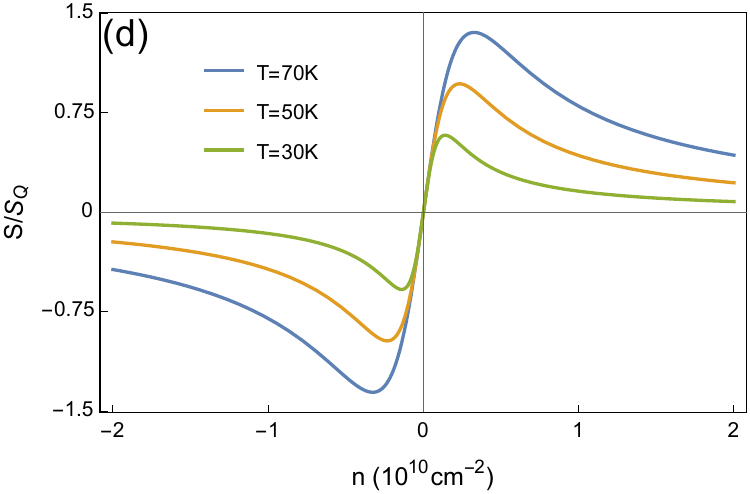}
\caption{Density dependence of transport coefficients for a graphene monolayer Corbino disk with the aspect ratio $r_2/r_1=3$ is illustrated for different temperatures: (a) the Lorenz ratio, next to the Wiedemann-Franz value  $L_{\text{WF}}=\pi^2/3e^2$ (dashed line), (b)  electric resistance normalized to $R_\sigma=\ln p/(2\pi\sigma)$, and (c) thermal conductance normalized to $G_T=2\pi T/\ln p$, (d) thermopower in units of $S_Q=\pi^2/3e$ (in units $\hbar=k_B=1$).  }
\label{Fig-LRGS}	
\end{figure*}

\subsection{Electrical resistance} 

To find the electric resistance $R$, we set the net temperature drop $\Delta T$ to zero and find that $R^{-1}$ is the $11$ matrix element  of $\hat{\mathcal{G}}$, i.e., $R^{-1}=\mathcal{G}_{11}=\mathcal{R}_{22}/\Det\hat{\mathcal{R}}$. Using the matrix elements of $\mathcal{\hat{R}}$ in Eq.~\eqref{R}, we obtain
\begin{equation}\label{R-el}
R^{-1}=\frac{2\pi e^2}{\ln p}\frac{\varsigma^2}{\varsigma+n\varkappa/s}+\frac{\pi e^2n^2r^2_2}{\eta(p^2-1)}.
\end{equation}
The second term here is inversely proportional to the shear viscosity of the liquid. It represents the contribution of the hydrodynamic transport mode to the electrical conductance. The first term is determined by the intrinsic transport coefficients of the electron liquid and represents the contribution of the relative transport mode to the conductance. The additivity of these contributions  to the conductance  is in stark contrast to Matthiessen's rule. Violation of the latter is one of the hallmarks of hydrodynamic transport.

For Galilean-invariant liquids, where $\varsigma=0$, the first term in Eq.~\eqref{R-el}  vanishes. However, in a generic conductor, the Galilean invariance is expected to be broken by the underlying crystalline lattice, and this term does not vanish. In particular, in graphene near charge neutrality, $n/s \ll 1$, this term is particularly pronounced. Precisely at charge neutrality, the second term vanishes and device resistance is determined by the intrinsic conductivity of the electron liquid, $R=\ln p/(2\pi\sigma)$. In contrast, at high density, $n/s \gg 1$, the viscous term prevails. Neglecting the first term, we recover the result of Ref. \cite{Shavit},  $R=\eta(r^{-2}_1-r^{-2}_2)/\pi(en)^2$.  However, even in the high density regime, the presence of the first term in
Eq.~\eqref{R-el} implies that the electric field does not vanish in the bulk flow in pristine systems with broken Galilean invariance. The appearance of the electric field in the bulk is caused by the temperature drop at the system boundary. Since the electrical conductance is measured at zero net temperature difference, the latter must be compensated by the temperature gradients in the bulk. Due to the force balance condition \eqref{eq:force_expulsion}, these gradients induce the EMF in the bulk.

\subsection{Lorenz ratio}

Let us now determine the Lorenz ratio $L=R/(TR_{\text{th}})$. In the Corbino geometry, it exhibits a very sensitive density dependence near charge neutrality. To determine this dependence, we focus on the entropy-dominated regime. Retaining the leading order terms $n/s \ll 1$ in Eq. \eqref{R}, we find
\begin{equation}\label{Lorentz}
L=\frac{1}{e^2}\left(\frac{s\Gamma}{n^2+\Gamma^2}\right)^2, \quad \Gamma^2=\frac{\sigma}{e^2}\frac{2\eta}{r^2_2}\left(\frac{p^2-1}{\ln p}\right).
\end{equation}
Since $\Gamma$ is inversely proportional to the system size, the width of the peak at charge neutrality is significantly smaller than $s$. Therefore, in the above expression for $\Gamma$, all quantities may be evaluated at charge neutrality.

It is apparent that at zero doping, $n\to0$, the Lorenz ratio may greatly exceed the Wiedemann-Franz value of $L_{\text{WF}}=\pi^2/3e^2$. For a graphene monolayer, their ratio may be estimated as $L/L_{\text{WF}}\propto (r_1/\lambda_T)^2$, where $\lambda_T\sim v/T$ is the thermal de Broglie wavelength.  For a typical micron size of the disk, one concludes that $L$ can be as high as $L/L_{\text{WF}}\gtrsim 10$ at temperatures $T>50$ K where electron hydrodynamic behavior is observed. This behavior is illustrated in Fig. \ref{Fig-LRGS}(a).   Furthermore, since the intrinsic conductivity is only weakly (logarithmically) temperature dependent, the width of the peak is primarily governed by the fluid viscosity $\eta$. For the Dirac liquid, it scales linearly with the temperature,  $\Gamma\propto T$.

\subsection{Thermoelectric effects} 

Lastly, we can determine the Seebeck coefficient $S=-(V/\Delta T)_{I=0}$ and Peltier coefficient $\Pi=(I_Q/I)_{\Delta T=0}$. They are  connected by the Onsager relation $S =\Pi/T = \mathcal{R}_{12} / T\mathcal{R}_{22}$. A direct calculation yields the thermopower in the form
\begin{equation}
\label{S}
S= \frac{1}{e}\frac{s n}{n^2+\Gamma^2},
\end{equation}
At high density, it reduces to the ratio of entropy density to charge density $S=s/(ne)$,  which is the value in the ideal hydrodynamic limit \cite{Aleiner,AKS}.  We note that the maximal thermopower, $S_{\mathrm{max}} = s/2 \Gamma$ is achieved at rather small densities, $n = \Gamma$, and can substantially exceed the prediction by the Mott formula in the single-particle picture of transport. In Fig. \ref{Fig-LRGS}(b)-(d), we illustrate the predicted density dependence of the transport coefficients for graphene monolayer devices at different temperatures.

For completeness, we consider two additional aspects of this transport problem. As any realistic sample has some degree of disorder, we include frictional forces in the analysis of hydrodynamic flow. This treatment is presented in Appendix \ref{app:disorder}, where we use the model of long-range disorder potential which is applicable to high mobility graphene samples. In Appendix \ref{app:ballistic}, we discuss electron transport in Corbino geometry in the ballistic regime, which may be realized in clean samples at low temperatures where the hydrodynamic description breaks down.

\section{Summary}

 We obtained the thermoelectric resistance matrix of Corbino devices in the hydrodynamic regime, Eq.~\eqref{R}. It is comprised from the contributions of two transport mechanisms.  The contribution of the  hydrodynamic transport mechanism is described by the first terms in the square brackets in Eq.~\eqref{R}. It corresponds to drops of temperature and voltage, which are localized at the sample boundaries, and arises from the dissipation caused by the viscous stresses in the bulk flow. Therefore, it cannot be written as a sum of two positive-definite contributions of the two boundaries. The contribution to the thermoelectric resistance  of charge and heat transport relative to the electron liquid is  described by the second terms in the square brackets in Eq.~\eqref{R}. It accounts for the voltage and temperature gradients in the bulk flow. In the absence of Galilean invariance, the electric field inside the liquid does not vanish in linear resistance measurements.

We applied these results to determine the thermal resistance, Eq.~\eqref{R-th}, electrical resistance, Eq.~\eqref{R-el}, the Lorenz ratio, Eq.~\eqref{Lorentz} and the Seebeck coefficient, Eq.~\eqref{S}. All transport coefficients exhibit a sensitive dependence on the electron density with the characteristic scale $n \sim \Gamma $, which is governed by the liquid viscosity and the sample size, see Eq.~\eqref{Lorentz}.  The hydrodynamic effects are manifested in strongly enhanced Lorenz ratio and thermoelectric power.

\begin{figure*}[t!]
\includegraphics[width=0.325\linewidth]{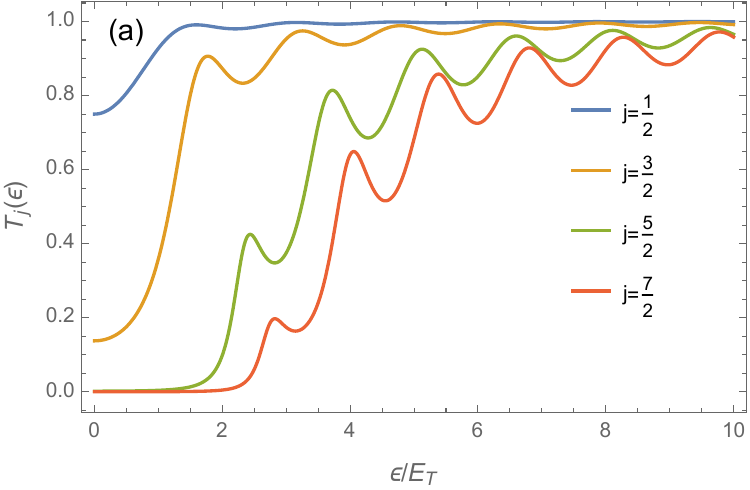}
\includegraphics[width=0.325\linewidth]{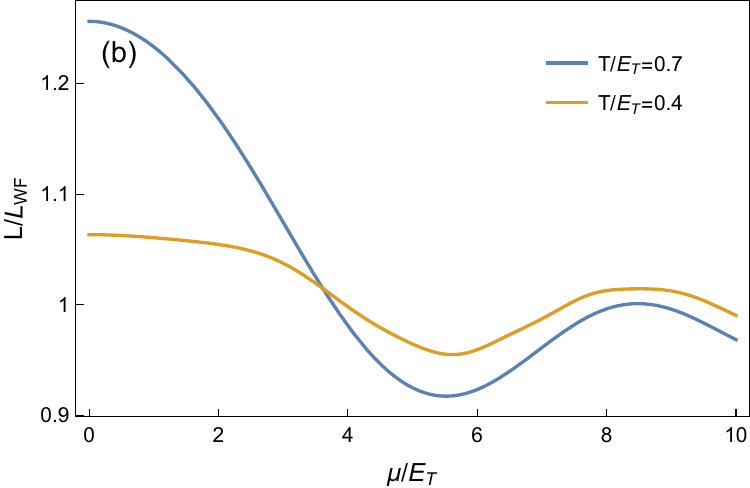}
\includegraphics[width=0.325\linewidth]{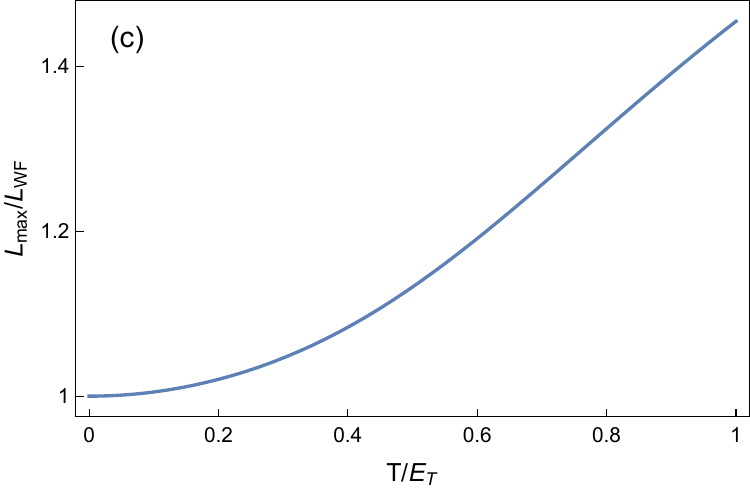}
\caption{The left panel-(a) shows energy dependence of partial transmission coefficients for different eigenmodes of quantization.  The central panel-(b) displays the Lorenz ratio for a Corbino disk as a function of the chemical potential as computed from Eq. \ref{Lorentz-ballistic} at two different temperatures of the ballistic regime $T<E_T$. The right panel-(c) shows temperature dependence of the normalized Lorenz ratio at the neutrality point $\mu\to0$.}
\label{Fig-TLL}	
\end{figure*}

\section*{Acknowledgments} We gratefully acknowledge illuminating discussions with Gregory Falkovich, Shahal Ilani, Philip Kim, Artem Talanov, and Jonah Waissman of various physical phenomena relevant to this work. 

This research was supported by the National Science Foundation Grant No. DMR-1653661 (S. L.), by the U.S. Department of Energy, Office of Science, Basic Energy Sciences Program for Materials and Chemistry Research in Quantum Information Science under Award No. DE-SC0020313 (A. L.), and by the MRSEC Grant No. DMR-1719797 (A. V. A). This project was initiated during the workshop ``From Chaos to Hydrodynamics in Quantum Matter" at the Aspen Center for Physics, which is supported by National Science Foundation Grant No. PHY-1607611.

\appendix

\section{Disorder effects}\label{app:disorder}

To establish a closer connection to realistic graphene Corbino devices, we discuss the impact of disorder scattering in the bulk of the flow. One of the main sources of disorder is believed to be due to charged impurities in the substrate on which graphene flake is deposited \cite{Crommie}. These impurities induce spatial fluctuations in the chemical potential, leading locally to regions of positive and negative charge density. This regime is commonly referred to as charge puddles. For boron nitride encapsulated graphene devices, scanning probes reveal that the correlation radius of these fluctuations is somewhere in the range $\xi\sim 100$ nm and local strength is in the range of $\sim 5$ meV \cite{STM-GhBn}. In the Corbino geometry, provided the length scale separation, $l\ll\xi\ll r_2$, one can average the flow of the electron fluid over the spatial inhomogeneities. This leads to an appearance of the effective friction force
\begin{equation}\label{friction}
\mathcal{F}=-ku,\quad k=\frac{\langle(s\delta n-n\delta s)^2\rangle}{2s^2}\frac{1}{\varsigma+n\varkappa/s},
\end{equation}
which needs to be added in the Navier-Stokes equation \eqref{NS}. This form of the friction coefficient $k$ was obtained in Ref. \cite{LLA}, where $\delta n(r)$ and $\delta s(r)$ denote local fluctuations of the particle and  entropy densities, respectively, and $\langle\ldots\rangle$ denotes spatial average. Accounting for $\mathcal{F}$ in Eq. \eqref{NS}, and repeating the same steps of derivation, it is easy to see that the special solution for $u(r)$ is now replaced by
\begin{equation}
u(r)=\frac{1}{2\pi r}\left(\frac{1}{1+kl^2/\eta}\right)\frac{{\vec x}^{\mathbb{T}}\hat{\Upsilon}^{-1}\vec{I}}{{\vec x}^{\mathbb{T}}\hat{\Upsilon}^{-1}\vec{x}}.
\end{equation}
As friction in part obstructs expulsion of the force from the bulk of the flow, we need to include an additional contribution to the energy dissipation of the form
\begin{equation}
\mathcal{P}_{\mathcal{F}}=k \int u^2(r) d^2r.
\end{equation}
When computing both terms in Eq. \eqref{P}, one finds that thermal resistance (at neutrality) is modified to
\begin{align}
R_{\text{th}}=\frac{2\eta(r^{-2}_1-r^{-2}_2)+k\ln p}{2\pi Ts^2}
\left(\frac{1}{1+kl^2/\eta}\right)^2\nonumber \\ 
+\frac{\ln p}{2\pi\kappa}\left(\frac{kl^2/\eta}{1+kl^2/\eta}\right)^2
\end{align}
We note here that in contrast to Eq. \eqref{R-th}, the bulk contribution no longer vanishes in the limit $n\to 0$. The form of the friction coefficient is also simplified at neutrality, where $k=(e^2/\sigma)\langle\delta n^2\rangle/2$. To estimate it, we notice that in the linear screening approximation the equilibrium density modulation is related to the external potential as $\delta n(q)=-\nu qU(q)/(q+a^{-1})$, where $a=1/(2\pi e^2\nu)$ is the Thomas-Fermi screening radius and $\nu\sim T/v^2$ is the thermodynamic single-particle density of states. In the hydrodynamic regime, correlation radius of disorder $\xi$ exceeds the Thomas-Fermi screening radius $a$. Therefore, $k\sim(e^2/\sigma)\langle U^2\rangle/(\xi^2e^4)$, where we assumed that the spectral density of disorder potential does not have strong divergence at $q\to0$ (e.g., encapsulation-induced disorder). Thus the friction term diminishes the contribution of the relative mode in the hydrodynamic regime since $kl^2/\eta\propto 1/T^4$, which decays rapidly with an increase of temperature. We also note that the temperature dependence of the friction contribution to resistance from $\mathcal{P}_\mathcal{F}$ decays faster with temperature than the viscous term since $k/(Ts^2)\propto 1/T^5$. The scattering off short-ranged quenched disorder gives an additional temperature independent contribution to the friction coefficient $k$. All other resistive coefficients can be modified accordingly.

\section{Ballistic regime}\label{app:ballistic}

For completeness, we briefly discuss thermoelectric matrix in the ballistic regime, which may be realized in clean systems at low temperatures, $T<E_T$, with $E_T=v/r_1$ being the characteristic Thouless energy of a Corbino disk. Adopting the Landauer-B\"{u}ttiker description \cite{Nazarov-Blanter}, all transport characteristics can be derived from the energy dependence of the transmission coefficient $\mathcal{T}(\varepsilon)$. For electrons traversing the monolayer graphene Corbino disk  the transmission coefficient can be computed analytically from the solution of the Dirac equation in cylindrical coordinates. It takes the form \cite{Rycrez,Katsnelson}
\begin{equation}
\mathcal{T}(\varepsilon)=\sum_{j}\mathcal{T}_j,\quad \mathcal{T}_j=\frac{16\lambda^2}{\pi^2r_1r_2}\frac{1}{\Gamma^2_+(\varepsilon)+\Gamma^2_-(\varepsilon)}
\end{equation}
where $\lambda=v/|\varepsilon|$ is the electron wavelength and the sum goes over the odd integers $j=n+1/2$ with $n\in\mathbb{Z}$. The functions in the denominator capture geometrical resonances and are given by
\begin{align}
\Gamma_\pm(\varepsilon)=\mathrm{Im}\left[\mathrm{H}^{(1)}_{j-1/2}(r_1/\lambda)\mathrm{H}^{(2)}_{j\mp1/2}(r_2/\lambda)\right. \nonumber \\
\left. \pm \mathrm{H}^{(1)}_{j+1/2}(r_1/\lambda)\mathrm{H}^{(2)}_{j\pm1/2}(r_2/\lambda)\right]
\end{align}
with $\mathrm{H}^{(1,2)}_n(z)$ the Hankel function of the (first, second) kind. For instance, in this formalism, the Lorenz ratio can be then computed as follows:
\begin{equation}\label{Lorentz-ballistic}
L=\frac{\mathcal{L}_0\mathcal{L}_2-\mathcal{L}^2_1}{e^2T^2\mathcal{L}^2_0},\quad
\mathcal{L}_n=\int\frac{d\varepsilon}{T}\frac{(\varepsilon-\mu)^n\mathcal{T}(\varepsilon)}{\cosh^2\left(\frac{\varepsilon-\mu}{2T}\right)}
\end{equation}
For the contrast to the results of hydrodynamic theory, we plot numerical results for Eq. \eqref{Lorentz-ballistic} in Fig. \ref{Fig-TLL}. The Lorenz ratio exhibits oscillations as a function of chemical potential that reflects geometrical resonances in the transmission coefficient. Exactly at neutrality, the Lorenz ratio for Dirac fermions stays above $L_{\text{WF}}$ and shows moderate growth with an increase of temperature. For $T/E_T>5$, the curve gradually saturates to the constant $L/L_{\text{WF}}\approx 2.37$. This regime is not shown in the plot as it is beyond the validity of single-particle description, since we expect a crossover to the collision-dominated regime to occur at $T\sim E_T$.

\end{document}